\documentclass[12pt,aps,prb]{revtex4}

\usepackage{amsmath}

\begin{document}
\title{A dipole in a dielectric: Intriguing results and shape
dependence of the
distant electric field }
\author{R. L. P. G. Amaral and N. A. Lemos}
\affiliation{Departamento
de F\'{\i}sica, Universidade Federal Fluminense,
Av. Litor\^anea s/n, Boa Viagem - CEP 24210-340,
Niter\'oi - Rio de Janeiro Brazil}

\begin{abstract}
The field of a point electric dipole in an
infinite dielectric is obtained by placing the dipole at the center
of a spherical cavity of radius $R$ inside the dielectric and then
letting $R\to 0$. The result disagrees with the elementary
answer found in textbooks. The mathematical and physical reasons
for the disagreement are discussed. The discrepancy is confirmed by
the same limiting procedure applied to a uniformly polarized sphere
embedded in the dielectric. We next solve the same problem
for a polarized spheroid immersed in an infinite dielectric and
find that the asymptotic potential shows an unexpected shape
dependence, even after taking the limit of an arbitrarily small
spheroid. By considering both oblate and prolate spheroids and
taking appropriate limits, we recover either the elementary
textbook answer or the previous result found for the polarized
sphere.
\end{abstract}

\maketitle

\section{Introduction}
Historically, electromagnetism, and particularly
electrostatics, has been a rich source of
beautiful mathematical physics problems, most of which are quite
standard by now. Yet, from time to time, a closer look at certain
simple and seemingly exhausted problems might surprise even the
experienced practitioner. We start by discussing the elementary
problem of determining the electrostatic field produced by a pure
(point) dipole embedded in the bulk of an infinite linear
dielectric medium. This problem is solved by two apparently
equivalent methods. The first makes use of an elementary argument
found in textbooks and the other consists of putting the dipole at
the center of a spherical hole in the dielectric and then letting
the radius of the hole tend to zero. The
discrepancy between the results might surprise the reader as much as
it surprised the authors. The discrepancy is corroborated by the
same limiting procedure applied to a uniformly polarized sphere
embedded in the dielectric.

Next we solve for the electrostatic field of a uniformly polarized
spheroid in an infinite dielectric. This solution is an interesting
exercise in mathematical physics involving in a simple way Legendre
functions of the second kind, which are seldom used in the standard
electromagnetism textbooks. We find that the asymptotic potential
exhibits a shape dependence. By taking appropriate limits, we
recover either the elementary textbook answer or the previous
result found for the polarized sphere. The dependence of the
electrostatic potential on the shape of the spheroid, even after
taking the limit in which the spheroid shrinks away keeping a
finite dipole moment, is unexpected and to a certain extent
non-intuitive. This physical effect
appears to have been overlooked by standard textbooks.

\section{Field of a Dipole in a Dielectric}

The problem of obtaining the field produced by a dipole in a
dielectric medium is one of those elementary problems that is
present (solved or proposed) in a variety of textbooks. The
well-known solution is trivial. The physical dipole consists of
two opposite point charges ($q$ and $-q$) separated by the distance
$d$. Letting $d\to 0$ with $qd=p_0$ gives the pure dipole. Because
in a linear dielectric medium Gauss' law $\oint{\bf
D}\cdot{\bf da}=q$ establishes that each of the charges $q$ and
$-q$ will be screened by polarization charges to
$q^{\prime}=q\epsilon_0/\epsilon$, the dipole moment will be
screened by the same factor, so that the actual (effective) dipole
moment is
\begin{equation}
\label{trivial}
p=p_0 \frac{\epsilon_0}{\epsilon}\,.
\end{equation}
Equation~(\ref{trivial}) is the answer found in standard
textbooks (see Ref.~\onlinecite{Griffiths} for example). In other
words, for a point dipole parallel to the $z$-axis and located at
the origin, the
electrostatic potential inside the infinite linear dielectric
medium in spherical coordinates
$(r,\theta,\varphi)$ is 
\begin{equation}
\label{dipolo}
\Phi({\bf r})=\frac{p}{4\pi\epsilon_0} \frac{\cos\theta}{r^2}\,,
\end{equation} with $p$ given by Eq.~(\ref{trivial}).

Now let us solve the ``same" problem by putting the pure dipole
${\bf p_0}$ at the center of an empty spherical hole of radius $R$
cut out of the dielectric medium and then letting $R\to 0$. It is
appropriate to make use of the general solution to Laplace's
equation in spherical coordinates for problems with azimuthal
symmetry. It is easy to see that the boundary conditions can be
satisfied by taking only the $\ell =1$ term of the azimuthally
symmetric general solution, so the electrostatic potential inside
the hole is
\begin{subequations}
\begin{equation}
\label{potinside}
\Phi^{(1)}({\bf r})=Ar\cos\theta + \frac{p_0}{4\pi\epsilon_0}
\frac{\cos\theta}{r^2}\,, \qquad (0<r<R)
\end{equation}
and the potential outside is
\begin{equation}
\label{potoutside}
\Phi^{(2)}({\bf r})=\frac{p^{\prime}}{4\pi\epsilon_0}
\frac{\cos\theta}{r^2}\,. \qquad (r>R)
\end{equation}
\end{subequations}

Note that inside the hole the singular term corresponds to the pure
dipole singularity with dipole moment
$p_0$, because the dipole is in vacuum. 
Outside, only the term that decreases with $r$ is present, with the
factor
$p^{\prime}$ to be determined. By requiring the continuity of the
scalar potential (equivalent to the continuity of the tangential
component of the electric field) and of the radial component of
the electric displacement vector (${\bf D}^{(1)}= -\epsilon_0
{\boldmath\nabla} \Phi^{(1)}$ and
${\bf D}^{(2)}= -\epsilon {\boldmath \nabla} \Phi^{(2)}$) at the
boundary
$r=R$, we obtain
\begin{eqnarray}
\label{Dcontinuo} -\epsilon_0 \frac{\partial \Phi^{(1)}}{\partial
r}\bigg\vert_{r=R} &=& -\epsilon \frac{\partial
\Phi^{(2)}}{\partial r}\bigg\vert_{r=R}\,,\\
\noalign{\noindent and}
\label{potcontinuo}
\Phi^{(1)}(R)&=&\Phi^{(2)}(R)\, .
\end{eqnarray}

The application of the boundary conditions in
Eqs.~(\ref{Dcontinuo}) and (\ref{potcontinuo}) leads to the
equations
\begin{eqnarray}
\label{Dcontinuo1}
 \epsilon_0\Bigl[\frac{2p^0}{4\pi\epsilon_0 R^3} - A\Bigr] &=&
\epsilon\,\frac{2p^{\prime}}{4\pi\epsilon_0 R^3} \,,\\
\noalign{\noindent and}
\label{potcontinuo1}
 \frac{p^0}{4\pi\epsilon_0 R^2} + A R &=&
\frac{p^{\prime}}{4\pi\epsilon_0 R^2}\,,
\end{eqnarray}
whose solution is
\begin{eqnarray}
\label{coeficientes1} A&=&\frac{2(\epsilon_0 -
\epsilon)}{2\epsilon +
\epsilon_0}\frac{p_0}{4\pi\epsilon_0R^3} \, \\
\noalign{\noindent and}
\label{dipololinha}
p^{\prime}&=&\frac{3\epsilon_0}{2\epsilon + \epsilon_0}p_0\, .
\end{eqnarray}
According to Eqs.~(\ref{coeficientes1}) and (\ref{dipololinha}),
the electrostatic potential outside the hole is that of a point
dipole in vacuum with effective dipole moment
$p^{\prime}$ given by Eq.~(\ref{dipololinha}).
In the limit $R\to 0$, the dipole potential everywhere except at
the origin is given by Eq.~(\ref{potoutside}) with $p^{\prime}$
determined by Eq.~(\ref{dipololinha}). Surprisingly, this effective
dipole moment disagrees with the one given in Eq.~(\ref{trivial})
by means of the previous elementary argument.

The reason for the discrepancy appears to be the lack of
commutativity of two successive limits. The result (\ref{trivial})
corresponds to putting the two opposite charges outside the hole in
the dielectric, letting the radius of the hole tend to zero first,
and then making the distance between the charges arbitrarily small,
thus creating a dipole at the origin. To obtain the result
(\ref{dipololinha}), we first let the distance between the charges
tend to zero, creating a point dipole at the center of the hole,
and only later do we make the radius of the hole arbitrarily small.
A physical explanation for the discrepancy is that in the first
case, but not in the second case, the charges are always screened
by the dielectric.

We might argue that the dipole moment associated with the
polarization charges on the surface of the hole added to $p_0$
leads to a total dipole moment given by Eq.~(\ref{dipololinha}),
which is in fact vindicated by an explicit calculation. This
argument, however, misses the point. The surprise comes from the
fact that, if only the {\it free} dipole moment $p_0$ is
considered, its reduction by the dielectric constant factor
does not account for the screening effect due to the polarization
of the medium. This behavior contrasts sharply with that of a point
charge at the center of the hole, whose field in the interior of
the dielectric is obtained by simply replacing the {\it free}
charge $q$ by
$q\epsilon_0 /\epsilon$ in the vacuum field.

\section{Uniformly Polarized Sphere in a
Dielectric}\label{sec:uniform}
To check the previous result in
Eq.~(\ref{dipololinha})
 and allow for a generalization in Sec.~\ref{uniform}, let
us consider a uniformly polarized sphere (electret) of radius $R$,
with polarization ${\bf P}$ along the $z$ axis, ${\bf P} = P_0 {\hat
{\bf k}}$, surrounded by an infinite dielectric whose dielectric
constant is
$\epsilon$. The potential has no singularity inside the sphere, so
we have
\begin{subequations}
\begin{equation}
\label{potinside1}
\Phi^{(1)}({\bf r})=Br\cos\theta \qquad (0<r<R)
\end{equation}
for the potential inside the sphere, while the
potential outside is
\begin{equation}
\label{potoutside1}
\Phi^{(2)}({\bf r})=\frac{p^{\prime}}{4\pi\epsilon_0}
\frac{\cos\theta}{r^2}\,. \qquad (r>R)
\end{equation}
\end{subequations}
We now notice that ${\bf D}^{(1)}= -\epsilon_0
\boldmath{\nabla}\Phi^{(1)} + {\bf P}$ and apply the
same boundary conditions as before to obtain
\begin{eqnarray}
\label{Dcontinuo2}
\frac{2\epsilon \, p^{\prime}}{4\pi\epsilon_0 R^3} &=&
-\epsilon_0 B + P_0 \,\\
\noalign{\noindent and}
\label{potcontinuo2}
\frac{p^{\prime}}{4\pi\epsilon_0 R^3} &=& B \,,
\end{eqnarray}
which are solved by
\begin{eqnarray}
B&=&\frac{P_0}{2\epsilon
+ \epsilon_0}\,, \\
\noalign{\noindent and}
\label{coeficientes2}
p^{\prime}&=& \frac{4\pi\epsilon_0}{2\epsilon
+\epsilon_0} R^3 P_0\,.
\end{eqnarray}
The resulting electrostatic potential inside the
polarized sphere is
\begin{subequations}
\begin{equation}
\label{potinside2}
\Phi^{(1)}({\bf r})= \frac{P_0}{2 \epsilon + \epsilon_0}
r\,\cos\theta\,, \qquad (0<r<R)
\end{equation}
and the potential outside is
\begin{equation}
\label{potoutside2}
\Phi^{(2)}({\bf r})=\frac{R^3 P_0}{2 \epsilon + \epsilon_0}
\frac{\cos\theta}{r^2}\,. \qquad (r>R)
\end{equation}
\end{subequations}
If we let $R \to 0$ and $P_0 \to \infty$ in such a
way that $p_0=(4/3)\pi R^3 P_0$ remains fixed, we would expect to
recover the point dipole ${\bf p}_0$ at the origin embedded in the
infinite dielectric. In such a limit, the potential everywhere
except at the origin becomes
\begin{equation}
\label{potoutside2a}
\Phi^{(2)}({\bf r})=\frac{3 \epsilon_0}{2 \epsilon +
\epsilon_0}\frac{1}{4\pi\epsilon_0}
\frac{p_0\,\cos\theta}{r^2}\,. \qquad (r>0)
\end{equation}
This result coincides with the $R \to 0$ limit of
the previous problem of the point dipole at the center of an empty
sphere inside the dielectric.

Here, again, the dipole moment of the polarization charges on the
spherical surface of the dielectric leads to the total dipole
moment (\ref{dipololinha}). Thus, the field inside the dielectric
is obtained from the vacuum field by reducing the free dipole
moment by a factor that differs from the screening factor for a
point charge.

\section{Uniformly Polarized Spheroid in a
Dielectric}\label{uniform}
To put the results of Sec.~\ref{sec:uniform} in a broader
context, which will make possible a further investigation of the
origin of the discrepancy encountered above, we will examine a
third ``interpolating" problem. Consider a uniformly polarized hole
(electret) in the dielectric medium with the shape of a spheroid (an
ellipsoid of revolution).

\subsection{The Oblate Case}
The oblate spheroidal coordinates are defined by (see
Ref.~\onlinecite{Arfken} for example)
\begin{equation}
\label{oblate}
\begin{array}{l} x=a\cosh \mu \sin v \cos \varphi \\
y=a\cosh \mu \sin v \sin \varphi \\
z=a\sinh \mu \cos v\,,
\end{array}
\end{equation}
with $\mu\geq 0$, $0\leq v\leq \pi$, $0\leq
\varphi\leq 2\pi$, and $a$ a positive real number. The surface of
the spheroid is defined by $\mu =\mu_0$, while its interior is
determined by $\mu < \mu_0$. It is easy to see that the surface of
the spheroid is given in cartesian coordinates by
\begin{equation}
\label{ellipsoid}
\frac{x^2}{X^2} + \frac{y^2}{X^2} + \frac{z^2}{Z^2} =1\,,
\end{equation}
where $X=a\cosh \mu_0$ and $Z=a\sinh \mu_0$, so that
$X>Z$. The ellipsoid is oblate, that is, flattened along the $z$
direction.

In terms of the new variables
\begin{equation}
\begin{array}{l}
\xi =\cos v \qquad (-1\leq \xi \leq 1)
\\
\zeta =\sinh \mu\,, \qquad (0\leq \zeta <
\infty)
\end{array}
\end{equation}
we can write
\begin{equation}
\label{oblate2}
\begin{array}{l} x=\rho \cos \varphi \\
y=\rho \sin
\varphi \\ z=a\xi \zeta\,,
\end{array}
\end{equation} with
\begin{equation}
\rho = a\bigl[ (1-\xi^2)(1+\zeta^2)\bigr]^{1/2}\, .
\end{equation}
The surface of the spheroid is now given by $\zeta =
\zeta_0$. Laplace's equation for the potential is separable in
these coordinates,\cite{Arfken} and its solution with rotational
symmetry about the $z$ axis, which is acceptable in the present
physical circumstances, is
\begin{subequations}
\begin{equation}
\Phi^{(1)}(\xi, \zeta) = \sum_{\ell =0}^{\infty} P_{\ell}(\xi)
\bigl[ A_{\ell} P_{\ell}(i\zeta) + B_{\ell} Q_{\ell}(i\zeta)
\bigr] \qquad (\zeta < \zeta_0)
\end{equation} inside the spheroid, and
\begin{equation}
\Phi^{(2)}(\xi, \zeta) = \sum_{\ell =0}^{\infty} P_{\ell}(\xi)
\bigl[ C_{\ell} P_{\ell}(i\zeta) + D_{\ell} Q_{\ell}(i\zeta)
\bigr] \qquad (\zeta > \zeta_0)
\end{equation}
\end{subequations}
outside the spheroid, where $P_{\ell}$ is the
$\ell$th Legendre polynomial and
$Q_{\ell}$ is the Legendre function of the second kind of order
$\ell$. The absence of
$Q_{\ell}(\xi)$ is necessary to guarantee the regularity
of $\Phi$ on the
$z$ axis ($\xi =1$).

An inspection of Eq.~(\ref{oblate}) shows that asymptotically
$\zeta$ plays the role of a radial coordinate. More precisely, for
large
$\mu$, we have $\xi\approx \cos\theta$ and $\zeta \approx
r/a$ with $r, \theta$ spherical coordinates. This observation
strongly suggests that the terms with $\ell=1$ alone will suffice
to satisfy the boundary conditions, and accordingly we take
\begin{subequations}
\begin{equation}
\Phi^{(1)}(\xi, \zeta) = P_1(\xi) \bigl[ A P_1(i\zeta) + B
Q_1(i\zeta) \bigr]\ \qquad (\zeta < \zeta_0)
\end{equation}
\begin{equation}
\Phi^{(2)}(\xi, \zeta) = P_1(\xi) \bigl[ C P_1(i\zeta) + D
Q_1(i\zeta) \bigr]\,, \qquad (\zeta > \zeta_0)
\end{equation}
\end{subequations}
where
\begin{equation}
P_1(\xi)=\xi \quad \mbox{and} \quad Q_1(i\zeta)= \zeta
\cot^{-1}\zeta -1
\, .
\end{equation}
It is not difficult to show that for large
$\zeta$
\begin{equation}
\label{asymptotic1} Q_1(i\zeta) \longrightarrow
-\frac{1}{3\zeta^2} \, .
\end{equation}
Therefore, the correct asymptotic behavior of $\Phi$
requires that $C=0$. As in the spherical coordinates case, it is
necessary to take
$B=0$ to avoid unphysical singularities. Indeed, the
$\xi$-component of the electric field associated with the term
$P_1(\xi)Q_1(i\zeta)$ is proportional to
$h_{\xi}^{-1}\partial[P_1(\xi)Q_1(i\zeta)]/\partial
\xi=a^{-1}(1-\xi^2)^{1/2}
(\xi^2+\zeta^2)^{-1/2}(\zeta\cot^{-1}\zeta -1)$, which is infinite
at $\xi =\zeta =0$, that is, at the circumference $\rho = a$ on the
$xy$-plane. Thus, we try to satisfy the boundary conditions with
(the imaginary unit has been absorbed into the coefficient $A$)
\begin{subequations}
\label{pot3}
\begin{equation}
\label{potinside3}
\Phi^{(1)}(\xi, \zeta) = A\xi\zeta\,, \qquad
(\zeta < \zeta_0)
\end{equation} and
\begin{equation}
\label{potoutside3}
\Phi^{(2)}(\xi, \zeta) = D\xi (\zeta \cot^{-1}\zeta -1)\,.
\qquad (\zeta > \zeta_0)
\end{equation}
\end{subequations}

The continuity of the potential at the surface of the spheroid
yields
\begin{equation} A\zeta_0 = D(\zeta_0 \cot^{-1}\zeta_0 -1)\,
.
\end{equation}
The continuity of the normal component of
$\bf D$ on the surface of the spheroid demands that
\begin{equation}
\label{Dcontinuo3} -\epsilon_0 \frac{1}{h_{\zeta}}\frac{\partial
\Phi^{(1)}}{\partial \zeta}\bigg\vert_{\zeta_0} + {\bf P}\cdot
{\hat{\bf e}}_{\zeta}\bigg\vert_{\zeta_0} = -\epsilon
\frac{1}{h_{\zeta}}\frac{\partial \Phi^{(2)}}{\partial
\zeta}\bigg\vert_{\zeta_0} \,,
\end{equation}
with
\begin{equation} h_{\zeta}= a\, \Bigl(\frac{\xi^2 +
\zeta^2}{1+\zeta^2}\Bigr)^{1/2}\,\,\, .
\end{equation} Because the unit outward normal vector to the
surface of the spheroid is
\begin{equation} {\hat{\bf e}}_{\zeta} = \frac{\partial {\bf
r}/\partial \zeta}{\vert \partial {\bf r}/\partial \zeta\vert} =
\frac{1}{h_{\zeta}}\frac{\partial {\bf r}}{\partial \zeta}\,,
\end{equation} it follows that
\begin{equation}
\label{Pnormal}
 {\bf P}\cdot {\hat{\bf e}}_{\zeta} = P_0 {\hat{\bf k}}\cdot
{\hat{\bf e}}_{\zeta}= P_0 \frac{1}{h_{\zeta}}{\hat{\bf k}}\cdot
\frac{\partial {\bf r}}{\partial \zeta} = P_0
\frac{a\xi}{h_{\zeta}}\, .
\end{equation}
The substitution of Eqs.~(\ref{pot3}) and (\ref{Pnormal}) into
Eq.~(\ref{Dcontinuo3}) leads to
\begin{equation} -\epsilon_0 A + P_0 a = -\epsilon D
\Biggl(\cot^{-1} \zeta_0 - \frac{\zeta_0}{1+\zeta_0^2}\Biggr)
\, .
\end{equation} The solution for $A$ and $D$ is
\begin{equation} A= \frac{(\cot^{-1}\zeta_0 -1/\zeta_0) a
P_0}{(\epsilon_0-\epsilon)\cot^{-1} \zeta_0 +
\frac{\epsilon\zeta_0}{1+\zeta_0^2} - \frac{\epsilon_0}{\zeta_0}}
\,,
\end{equation}
\begin{equation} D= \frac{a P_0}{(\epsilon_0-\epsilon)\cot^{-1}
\zeta_0 + \frac{\epsilon\zeta_0}{1+\zeta_0^2} -
\frac{\epsilon_0}{\zeta_0}}
\, .
\end{equation}

The volume of the spheroid is
\begin{equation}
\label{volume} V=\frac{4}{3}\pi (a\cosh \mu_0)^2 a\sinh \mu_0 =
\frac{4}{3}\pi a^3 (1+\zeta_0^2) \zeta_0
\,,
\end{equation} so that the dipole moment of the spheroid is
\begin{equation}
\label{dipolototal} p_0= \frac{4}{3}\pi a^3 (1+\zeta_0^2) \zeta_0\, P_0
\, .
\end{equation}

For large $\zeta$, we have $\xi \approx \cos\theta$ and
$\zeta \approx r/a$, so that, with the use of
Eqs.~(\ref{asymptotic1}) and (\ref{dipolototal}), we find that the
asymptotic behavior of the potential is
\begin{equation}
\label{potoblato}
\Phi^{(\mbox{ob})} (\xi, \zeta) \longrightarrow
\frac{\epsilon_0}{(\epsilon -
\epsilon_0)\zeta_0(1+\zeta_0^2)\cot^{-1}\zeta_0 + \epsilon_0 +
(\epsilon_0 - \epsilon) \zeta_0^2}\,
\frac{p_0}{4\pi \epsilon_0}\,\frac{\cos\theta}{r^2}
\, .
\end{equation}

This result is quite unexpected! Our intuition leads us to believe
that seen from far away, it is impossible to tell a uniformly
polarized sphere from a uniformly polarized ellipsoid. The shape
independence of the asymptotic potential prevails only for the
vacuum ($\epsilon=\epsilon_0$). For a dielectric, the
asymptotic potential depends on the shape of the dipole
distribution near the origin.

\subsection{The Prolate Case}

The analysis of the prolate case runs along similar lines. The
prolate spheroidal coordinates are defined by{\cite{Arfken}
\begin{equation}
\label{prolate}
\begin{array}{l} x=a\sinh \mu \sin v \cos \varphi \\
y=a\sinh \mu \sin v \sin \varphi \\
z=a\cosh \mu \cos v
\, .
\end{array}
\end{equation}
The surface of the spheroid is defined by $\mu
=\mu_0$, while its interior is determined by $\mu < \mu_0$. The
surface of the spheroid is again given in cartesian coordinates by
Eq.~(\ref{ellipsoid}) with $X=a\sinh \mu_0$ and $Z=a\cosh \mu_0$, so
that $Z>X$ and the spheroid is elongated in the $z$ direction.

In terms of the new variables
\begin{equation}
\begin{array}{l}
\xi =\cos v\ \qquad (-1\leq \xi \leq
1) \\
\eta =\cosh \mu\,, \qquad (1\leq \eta \leq
\infty)
\end{array}
\end{equation}
we can write
\begin{equation}
\label{prolate2}
\begin{array}{l} x=\rho \cos \varphi\ \\ y=\rho \sin
\varphi \\ z=a\xi \eta \,,
\end{array}
\end{equation} with
\begin{equation}
\rho = a\bigl[ (1-\xi^2)(\eta^2 - 1)\bigr]^{1/2} \, .
\end{equation} The surface of the spheroid is now given by $\eta =
\eta_0$, and the solution to Laplace's equation that suits our
problem is
\begin{equation}
\label{potinside4}
\Phi^{(1)}(\xi, \eta) = F P_1(\xi)P_1(\eta)\,,
\qquad (\eta < \eta_0)
\end{equation} and
\begin{equation}
\label{potoutside4}
\Phi^{(2)}(\xi, \eta) = G P_1(\xi)Q_1(\eta)\,.
\qquad (\eta > \eta_0)\, .
\end{equation} Here
\begin{equation} Q_1(\eta) = \frac{\eta}{2}\, \ln \frac{\eta +
1}{\eta - 1} - 1 \,,
\end{equation} whose asymptotic behavior for large $\eta$ is
\begin{equation}
\label{asymptotic2} Q_1(\eta) \longrightarrow \frac{1}{3\eta^2}
\, .
\end{equation}

The application of the boundary conditions at the surface of the
uniformly polarized prolate spheroid yields
\begin{equation} F= \frac{\bigg(\frac{1}{2}\, \ln \frac{\eta_0 +
1}{\eta_0 - 1} - \frac{1}{\eta_0}\bigg) a P_0}
{\frac{\epsilon_0-\epsilon}{2}\, \ln \frac{\eta_0 + 1}{\eta_0 - 1}
+ \frac{\epsilon\eta_0}{\eta_0^2 - 1} -
\frac{\epsilon_0}{\eta_0}}
\end{equation}
\begin{equation} G= \frac{a P_0}{\frac{\epsilon_0-\epsilon}{2}\,
\ln \frac{\eta_0 + 1}{\eta_0 - 1} + \frac{\epsilon\eta_0}{\eta_0^2
- 1} - \frac{\epsilon_0}{\eta_0}}
\, .
\end{equation}

The volume of the prolate spheroid is
\begin{equation}
\label{volume2} V=\frac{4}{3}\pi (a\sinh \mu_0)^2 a\cosh \mu_0 =
\frac{4}{3}\pi a^3 (\eta_0^2 - 1) \eta_0
\,,
\end{equation} and its dipole moment is
\begin{equation} p_0= \frac{4}{3}\pi a^3 (\eta_0^2 - 1) \eta_0\, P_0
\, .
\end{equation}

As for the oblate case, we have for large $\eta$, $\xi
\approx
\cos\theta$ and $\eta \approx r/a$, so that the asymptotic
behavior of the potential is
\begin{equation}
\label{potprolato}
\Phi^{(\mbox{pr})} (\xi, \zeta) \longrightarrow
\frac{\epsilon_0}{\frac{\epsilon_0 - \epsilon}{2} \eta_0(\eta_0^2 -
1)\ln \frac{\eta_0 + 1}{\eta_0 - 1}
 + \epsilon_0 + (\epsilon -\epsilon_0) \eta_0^2}\,
\frac{p_0}{4\pi \epsilon_0}\,\frac{\cos\theta}{r^2}
\, .
\end{equation}

Once again the asymptotic potential exhibits a surprising shape
dependence that is absent only in the case of the vacuum. For a
dielectric the asymptotic potential allows us to tell the
difference between a uniform dipole density distributed within a
sphere, an oblate spheroid, or a prolate spheroid. If a hole in the
dielectric is filled with a uniform charge density, no such shape
dependence is observed. The total induced charge depends only on
the internal free charge and on the dielectric constant.

Note also the striking result that the electric field is uniform
inside the spheroid because the potential is of the form
$\Phi^{(1)}= Az$ in both cases, as Eqs.~(\ref{potinside3}) and
(\ref{potinside4}) show.

\section{Limiting Cases and Conclusion}

Let us define the screening factor $\alpha$ as the coefficient that
multiplies the vacuum asymptotic dipole field to give the
asymptotic dipole field in the presence of the dielectric medium.
From Eq.~(\ref{potprolato}) it follows immediately that
\begin{equation}
\label{screenfactor}
\alpha = \epsilon_0\,\bigg[\frac{\epsilon_0 - \epsilon}{2}
\eta_0(\eta_0^2 - 1)\ln \frac{\eta_0 + 1}{\eta_0 - 1} 
 + \epsilon_0 + (\epsilon -\epsilon_0) \eta_0^2\bigg]^{-1}
\, .
\end{equation}

The spherical limit is reached by letting $\eta_0\equiv\cosh
\mu_0\to \infty$ and $a\to 0$ in such a way that $a\cosh \mu_0 =
R$ remains fixed. Then we obtain $X=Z=R$ in 
Eq.~(\ref{ellipsoid}) and the spheroid degenerates into a sphere.
If we take into account that for large $\eta_0$
\begin{equation}
\label{screenfactor2}
\ln \frac{\eta_0 + 1}{\eta_0 - 1} = \frac{2}{\eta_0} +
\frac{2}{3\eta_0^3} + \cdots\,,
\end{equation}
it is easy to show that
\begin{equation}
\label{screenfactor3}
\lim_{\eta_0 \to \infty} \alpha = \frac{3\epsilon_0}{2\epsilon +
\epsilon_0}
\, .
\end{equation} Thus our previous result for the uniformly polarized
sphere is recovered.

Let us now examine the line dipole limit, reached by letting
$\eta_0\to 1$ or, equivalently, $\mu_0\to 0$. In this
limit we have $X=0$ and $Z=a$, so that the ellipsoid
(\ref{ellipsoid}) reduces to a line segment (a rod) along the
$z$-axis. If we recall that $\lim_{x\to 0} x\ln x =0$, we
can readily show that 
\begin{equation}
\label{screenfactor4}
\lim_{\eta_0 \to 1} \alpha = \frac{\epsilon_0}{\epsilon}
\,,
\end{equation} and the standard answer (\ref{trivial}) is regained.

For the oblate spheroid we can also consider two limiting cases
with the help of Eq.~(\ref{potoblato}). If we let $\zeta_0 \to
\infty$ and $a\to 0$ with $a\zeta_0=R$, the spherical limit is
reached and $\alpha \to 3\epsilon_0/(2\epsilon + \epsilon_0)$, as
it should. If $\zeta_0 \to 0$, the spheroid becomes a ``pancake"
describing a dipole layer, and $\alpha \to 1$. This result appears
to be of some interest, inasmuch as the presence of the dielectric
does not change the vacuum field.
 
We believe that further discussion of the physical grounds for the
discrepancies is necessary. The screening factor $\alpha$ is a
measure of the total dipole moment in the presence of the dieletric
relative to the vacuum dipole moment $p_0$. Thus, the dipole
moment of the charges induced on the surface of the hole depends
not only on the free dipole moment $p_0$ and the dielectric
constant, but also on the shape of the hole. That the shape
dependence persists in the infinitely small hole limit seems to be 
related to the singularity of the dipole field, which is stronger
than that of the monopole field. For a shrinking finite
charge distribution (monopole), Gauss' law forbids this effect.
The reader might want to generalize the dipole result for
higher multipole moments.

It is suspected that such a shape dependence would manifest
itself in the dynamical case, that is, in the radiation from a
point dipole embedded in an infinite dielectric. Such a phenomenon
might be of relevance in condensed matter physics. For instance,
it might give rise to classical effects in the theory of quantum
dots. 
\vskip 1cm

\begin{acknowledgements}

This work was partially supported by Conselho Nacional de
Desenvolvimento Cient\'{\i}fico e Tecnol\'ogico (CNPq), Brazil.
Special thanks are due to David Griffiths, whose corrections and
suggestions have contributed to a significant improvement of the
paper.
\end{acknowledgements}


\newpage

\begin{thebibliography}{60}

\bibitem{Griffiths} D. J. Griffiths, {\it Introduction to
Electrodynamics} (Prentice Hall, New Jersey, 1999); see Problem
4.34 on p. 198 and let $R\to\infty$.

\bibitem{Arfken} G. Arfken, {\it Mathematical Methods for
Physicists} (Academic, New York, 1970), 2nd ed.

\end{thebibliography}
\end{document}